\documentstyle[12pt]{article}

\let\ni=\noindent

\begin{document}

\baselineskip 0.75cm
 
\pagestyle {empty}

\renewcommand{\thefootnote}{\fnsymbol{footnote}}

\newcommand{\CKM}{Cabibbo---Kobayashi---Maskawa }

\newcommand{\SK}{Super--\-Kamio\-kande }

~~~

\vspace{2.0cm}

{\large\centerline{\bf Two hypothetic sterile neutrinos which want to mix}}

{\large\centerline{\bf with $\nu_e $ and $\nu_\mu $}}

\vspace{0.8cm}

{\centerline {\sc Wojciech Kr\'{o}likowski}}

\vspace{0.8cm}

{\centerline {\it Institute of Theoretical Physics, Warsaw University}}

{\centerline {\it Ho\.{z}a 69,~~PL--00--681 Warszawa, ~Poland}}

\vspace{1.0cm}

{\centerline {\bf Abstract}}

\vspace{0.3cm}

 It is argued that the observed deficit of solar and atmospheric neutrinos can 
be explained by neutrino oscillations $\nu_e \rightarrow \nu_s $ and $\nu_\mu
\rightarrow \nu'_{s}$ involving two hypothetic sterile neutrinos $\nu_s $ and 
$\nu'_{s} $ (blind to all Standard--Model interactions). They are keen to mix 
nearly maximally with $\nu_{e}$ and $\nu_{\mu}$, respectively, to form neutrino
mass states $\nu_1$, $\nu_4 $ and $\nu_2 $, $\nu_5 $. Our argument is presented
in the framework of a model of fermion "texture" formulated previously, which 
implies the existence of two sterile neutrinos beside the three conventional.

\vspace{0.3cm} 

\ni PACS numbers: 12.15.Ff , 12.90.+b , 14.60.Gh 
 
\vspace{4.0cm} 

\ni July 1998

\vfill\eject

\pagestyle {plain}

\setcounter{page}{1}

~~~

\vspace{0.6cm}

\ni {\bf 1. Introduction}

\vspace{0.4cm}

 The recent findings~~[1]~~of \SK~~atmospheric--neutrino~~experiment brought to
us the important message that the observed deficit of atmospheric $\nu_\mu$'s 
seems to be really caused by neutrino oscil\-lations, related to nearly maximal
mixing of $\nu_\mu $ with another neutrino. This may be $\nu_\tau $ or, alter%
natively, a new sterile neutrino (blind to all Standard--Model interactions). 
The $\nu_e $ neutrino is here excluded from being a mixing partner of $\nu_\mu 
$ by the negative result of CHOOZ long--baseline reactor experiment [2] which 
found no evidence for the disappearance modes of $\bar{\nu}_e$, in particular 
$\bar{\nu}_e \rightarrow \bar{\nu}_\mu $, in a parameter region overlapping 
the range of $\sin^2 2\theta_{\rm atm}$ and $\Delta m^2_{\rm atm}$ observed in 
the \SK experiment.

 The survival probability for $\nu_\mu $, when analized experimentally in 
two--flavor form

%\vspace{-0.3cm}

%rownanie 1
\begin{equation}
P\left(\nu_\mu \rightarrow \nu_\mu\right) = 1 - \sin^2 2\theta_{\rm atm}
\sin^2 \left(1.27 \Delta m^2_{\rm atm}\, L/E \right)\;,
\end{equation}

\vspace{0.1cm}

\ni leads to the parameters [1]

%\vspace{-0.2cm}

%rownanie 2
\begin{equation}
\sin^2 2\theta_{\rm atm} = O(1) \sim 0.82\;\;{\rm to}\;\; 1
\end{equation}

\vspace{0.1cm}

\ni and

%\vspace{-0.2cm}

%rownanie 3
\begin{equation}
\Delta m^2_{\rm atm} \sim (0.5\;\;{\rm to}\;\; 6)\times 10^{-3}\;\;{\rm eV}^2
\end{equation}

\ni at the 90\% confidence level (note that the value $\Delta m^2_{\rm atm} 
\sim 5\times 10^{-3}\;\;{\rm eV}^2 $ corresponds to the lower limit of the 
previous Kamiokande estimate of $\Delta m^2_{\rm atm} $ [3]). If $\nu_\tau $ is
responsible for this nearly maximal mixing of $\nu_\mu$, then the disappearance
probability for $\nu_\mu $ in the mode $\nu_\mu \rightarrow \nu_\tau $ is

%\vspace{-0.2cm}

%rownanie 4
\begin{equation}
P\left(\nu_\mu \rightarrow \nu_\tau\right) = \sin^2 2\theta_{\rm atm}
\sin^2 \left(1.27 \Delta m^2_{\rm atm}\, L/E \right)\;.
\end{equation}

 In the present paper, we conjecture that it is rather a sterile neutrino 
(denoted here by $\nu'_s $) which is responsible for such a nearly maximal
mixing of $\nu_\mu $(whether it is not or is $\nu_\tau $ constitutes a crucial 
point of our conjecture which, unfortunately, is not at the moment easy to 
decide experimentally [1]). We conjecture moreover that another sterile 
neutrino (denoted by $\nu_s $) mixes nearly maximally with $\nu_e $, causing 
the observed deficit of solar $\nu_e $'s. In such a way, we introduce a unified
picture of neutrino oscillations as being related to nearly maximal mixing of 
two sterile neutrinos $\nu_s $ and $\nu'_s $ with $\nu_e $ and $\nu_\mu $, 
respectively. Of course, this mixing of $\nu_s $ and $\nu'_s $ is not forbidden
by the weak isospin $ I_3 $ and weak hypercharge $ Y $ of $\nu_e $ and $\nu_\mu
$ , as the conservation of these weak charges is spontaneously broken, except 
for their combination $ Q \equiv I_3 + Y/2 $ (equal to zero for $\nu_e $ and 
$\nu_\mu $). We should like also to remark that the sterile neutrinos $\nu_s $ 
and $\nu'_s $, interacting only gravitionally, would be responsible for the 
existence of a Standard--Model--inactive fraction of the dark matter.

 Note that the existence of just two sterile neutrinos (blind to all Standard%
--Model interactions), beside three families of Standard--Model--active leptons
and quarks, turns out to be natural in the model of lepton and quark "texture" 
we develop since some time [4,5] ({\it cf.} Eqs. (A.15) in Appendix). In this 
model, all neutrinos are Dirac particles having both lefthanded and right\-%
handed parts.

 For the Standard--Model--active neutrinos $\nu_e\,,\;\nu_\mu\,,\;\nu_\tau $,
charged leptons $ e^-\,,\;\mu^-\,,\;\tau^- $, up quarks $ u\,,\;c\,,\;t $ and 
down quarks $ d\,,\;s\,,\;b $ we came to a proposal [5] ({\it cf.} Eq. (A.10) 
in Appendix) of unified algebraic structure of their mass matrices $\left( 
M^{(f)}_{ij} \right)\;\, (f = \nu\,,\; e\,,\; u\,,\; d)$ in the three--dimen\-%
sional family space $ (i\,,\; j = 1\,,\, 2\,,\,3)$. In the case of leptons $(f 
= \nu\,,\; e)$, this proposal reads

% rownanie 5
\begin{equation}
\left({M}^{(f)}_{ij}\right) = \frac{1}{29} \left(\begin{array}{ccc} 
\mu^{(f)}\varepsilon^{(f)\,2} & 2\alpha^{(f)} e^{i\varphi^{(f)}} & 0 \\ & & 
\\ 2\alpha^{(f)} e^{-i\varphi^{(f)}} & 4\mu^{(f)}(80 + \varepsilon^{(f)\,2})/9 
& 8\sqrt{3}\alpha^{(f)} e^{i\varphi^{(f)}} \\ & & \\ 0 & 8\sqrt{3}\alpha^{(f)} 
e^{-i\varphi^{(f)}} & 24\mu^{(f)} (624 + \varepsilon^{(f)\,2})/25 \end{array}
\right) \;.
\end{equation}

\ni Here, $\mu^{(f)}$, $\varepsilon^{(f)\,2}$, $\alpha^{(f)}$ and $\varphi^{
(f)}$ denote real constants to be determined from the present and future exper%
imental data for lepton masses and mixing parameters ($\mu^{(f)}$ and $\alpha^{
(f)}$ are mass--dimensional).

 For charged leptons, when assuming that  the off--diagonal elements of the 
mass matrix $ \left({M}^{(e)}_{ij}\right)$ given in Eq. (5) can be treated as 
a small perturbation of its diagonal terms, we calculate in the lowest 
(quadratic) perturbative order in  $\alpha^{(e)}/\mu^{(e)}$ [5]:

\vspace{0.3cm}

%rownanie 6
\begin{eqnarray}
m_\tau & = & \frac{6}{125} \left(351 m_\mu - 136 m_e \right) \nonumber \\
& + & \frac{216 \mu^{(e)}}{3625} \left( \frac{111550}{31696 + 29
\varepsilon^{(e)\,2}} - \frac{487}{320 - 5\varepsilon^{(e)\,2}}\right)\,
\left(\frac{\alpha^{(e)}}{\mu^{(e)}}\right)^2 \;,\nonumber \\
\varepsilon^{(e)\,2} & = & \frac{320 m_e}{9 m_\mu - 4 m_e} + O\left[\left(
\frac{\alpha^{(e)}}{\mu^{(e)}}\right)^2 \right] \;, \nonumber \\
\mu^{(e)} & = & \frac{29}{320} \left(9 m_\mu - 4 m_e \right) + O\left[\left(
\frac{\alpha^{(e)}}{\mu^{(e)}}\right)^2\right]\mu^{(e)} \;.
\end{eqnarray}

\vspace{0.2cm}

\ni When the experimental $ m_e $ and $ m_\mu $ [6] are used as inputs, Eqs. 
(6) give [5]

\vspace{0.1cm}

%rownanie 7
\begin{eqnarray}
m_\tau & = & \left[ 1776.80 + 10.2112 \left(\frac{\alpha^{(e)}}{\mu^{(e)}}
\right)^2\,\right]\;{\rm MeV} \;, \nonumber \\
\varepsilon^{(e)\,2} & = & 0.172329 + O\left[\left(\frac{\alpha^{(e)}}{\mu^{(e)
}}\right)^2 \right] \;,\nonumber \\
\mu^{(e)} & = & 85.9924\;{\rm MeV} + O\left[\left(\frac{\alpha^{(e)}}{\mu^{(e)}
}\right)^2\,\right]\,\mu^{(e)}\;. 
\end{eqnarray}

\vspace{0.2cm}

\ni We can see that the predicted value of $ m_\tau $ agrees very well with its
experimental figure $ m_\tau^{\rm exp} = 1777.00^{+0.30}_{-0.27}$~MeV [6], even
in the zero--order perturbative calculation. To estimate $\left(\alpha^{(e)}/
\mu^{(e)}\right)^2 $, we take this experimental figure as another input. Then,

\vspace{0.1cm}

%rownanie 8
\begin{equation}
\left(\frac{\alpha^{(e)}}{\mu^{(e)}}\right)^2 = 0.020^{+0.029}_{-0.020} \;,
\end{equation}

\vspace{0.2cm}

\ni so it is not inconsistent with zero.

 The unitary matrix $\left({U}^{(e)}_{ij}\right)$, diagonalizing the mass 
matrix $\left({M}^{(e)}_{ij}\right)$ according to the relation $ U^{(e)\,
\dagger}\,M^{(e)}\,U^{(e)} = {\rm diag}(m_e\,,\,m_\mu\,,\,m_\tau)$, assumes in 
the lowest (quadratic) perturbative order in $\alpha^{(e)}/\mu^{(e)}$ the form

\vspace{0.3cm}

%rownanie 9
\begin{equation}
\left( U^{(e)}_{ij}\right) = \left(\begin{array}{ccc} 1 - \frac{2}{841}\left( 
\frac{\alpha^{(e)}}{m_\mu}\right)^2 & \frac{2}{29}\frac{\alpha^{(e)}}{m_\mu}
e^{i\varphi^{(e)}} & 0 \\ -\frac{2}{29}\frac{\alpha^{(e)}}{m_\mu}e^{-i\varphi^{
(e)}} & 1 - \frac{2}{841}\left(\frac{\alpha^{(e)}}{m_\mu}\right)^2 - \frac{96}{
841}\left(\frac{\alpha^{(e)}}{m_\tau}\right)^2 & \frac{8\sqrt{3}}{29}
\frac{\alpha^{(e)}}{m_\tau} e^{i\varphi^{(e)}} \\ 0 & -\frac{8\sqrt{3}}{29}
\frac{\alpha^{(e)}}{m_\tau}e^{-i\varphi^{(e)}} & 1  - \frac{96}{841}\left(
\frac{\alpha^{(e)}}{m_\tau}\right)^2 \end{array} \right) \;,
\end{equation}

%\vspace{0.2cm}

\ni where the small $\varepsilon^{(e)\,2}$ is neglected. Of course, in the 
limit of $\alpha^{(e)} \rightarrow 0$, we obtain $\left(U_{ij}^{(e)} \right)
\rightarrow \left(\delta_{ij}\right)$.

 For neutrinos, we will assume in this paper that $\varepsilon^{(\nu)\,2}$ 
is very small and

\vspace{-0.3cm}

%rownanie 10
\begin{equation}
\alpha^{(\nu)} = 0\;,
\end{equation}

\vspace{-0.15cm}

\ni in contrast to the possibility of $\alpha^{(e)} \neq 0 $ for charged 
leptons [{\it cf.} Eq. (8)]. Then, for conventional neutrinos $\left(U_{ij}^{(
\nu)} \right) = \left(\delta_{ij}\right)$ and so, $\nu_e$, $\nu_\mu $, $
\nu_\tau$ can mix only by means of the trivial lepton CKM matrix $ \left(V_{ij}
\right) \equiv \left(\sum_k U_{ki}^{(\nu)\,*}U_{kj}^{(e)}\right) = \left(U_{ij
}^{(e)}\right)$, what is a minor effect, vanishing in the limit of $\alpha^{
(e)} \rightarrow 0 $. Instead, allowing in this paper for the existence of two 
sterile neutrinos $\nu_s $ and $\nu'_s $, we will extend the $ 3\times 3 $ 
neutrino mass matrix $ \left({M}^{(\nu)}_{ij}\right)$ ($ i,j = 1,2,3 $), given 
through Eqs. (5) and (10), to a $ 5\times 5 $ neutrino mass matrix $\left( M^{(
\nu)}_{IJ}\right)$ ($I,J = 1,2,3,4,5$) with $ M^{(\nu)}_{IJ} = M^{(\nu)\,*}_{J
I}$. Explicitly, we will assume that

\vspace{-0.2cm}

% rownanie 11
\begin{equation}
\left({M}^{(\nu)}_{IJ}\right) = \left(\begin{array}{ccccc}M_{11}^{(\nu)} & 
0 & 0 & M_{14}^{(\nu)} & 0 \\ 0 & M_{22}^{(\nu)} & 0 & 0 & M_{25}^{(\nu)} \\ 
0 & 0 & M_{33}^{(\nu)} & 0 & 0 \\ M_{41}^{(\nu)} & 0 & 0 & M_{44}^{(\nu)} & 0
\\ 0 & M_{52}^{(\nu)} & 0 & 0 & M_{55}^{(\nu)}\end{array}\right) \,,
\end{equation}

\vspace{-0.1cm}

\ni where $ M^{(\nu)}_{11} = \mu^{(\nu)} \varepsilon^{(\nu)\,2}/29 $, $ M^{(\nu
)}_{22} \simeq 320 \mu^{(\nu)}/261 $, $ M^{(\nu)}_{33} \simeq 14976\mu^{(\nu)}/
725 $ due to Eq. (5), and $ M^{(\nu)}_{44} \sim  \mu^{(\nu)}\varepsilon^{(\nu)\,
2}/7 $, $ M^{(\nu)}_{55} \sim 48 \mu^{(\nu)}/7 $ in consequence of Eqs. (A.19) 
and (A.20) ({\it cf.} Appendix). It will turn out that the matrix elements $ 
M^{(\nu)}_{14} =  M^{(\nu)\,*}_{41} $ and $ M^{(\nu)}_{25} =  M^{(\nu)\,*}_{52}
$ lead to the mixing of neutrino flavor states $\nu_e $ with $\nu_s $ and $
\nu_\mu $ with $\nu'_s $ within neutrino mass states $\nu_1 $, $\nu_4 $ and $
\nu_2 $, $\nu_5 $, respectively.

\vspace{0.3cm}

\ni {\bf 2. Neutrino mass states}

\vspace{0.3cm}

 The eigenvalues of the extended mass matrix $\left({M}^{(\nu)}_{IJ}\right)$ 
given in Eq. (11) are Dirac masses of five neutrino mass states $\nu_1\,,\,
\nu_2\,,\,\nu_3\,,\,\nu_4\,,\,\nu_5 $. They are 

\vspace{-0.2cm}

%rownanie 12
\begin{eqnarray}
m_{\nu_1,\,\nu_4} & = & \frac{M^{(\nu)}_{11}+M^{(\nu)}_{44}}{2} \mp \sqrt{
\left(\frac{M^{(\nu)}_{11}-M^{(\nu)}_{44}}{2}\right)^2 + |M^{(\nu)}_{14}|^2}
\;, \nonumber \\ m_{\nu_3\,}\;\;\;\, & = & M^{(\nu)}_{33}\;, \nonumber \\
m_{\nu_2,\,\nu_5} & = & \frac{M^{(\nu)}_{22}+M^{(\nu)}_{55}}{2} \mp \sqrt{
\left(\frac{M^{(\nu)}_{22}-M^{(\nu)}_{55}}{2}\right)^2 + |M^{(\nu)}_{25}|^2}
\;.
\end{eqnarray}

\vspace{-0.1cm}

\ni In Section 4, the masses $ m_{\nu_1}$ and $ m_{\nu_2}$ will turn out to be 
negative, what is irrelevant in the case of Dirac particles for which only 
masses squared are measurable (so,  $ |m_{\nu_1}|$ and $ |m_{\nu_2}|$ will be 
the phenomenological masses of $\nu_1 $ and $\nu_2 $).

 The corresponding $5\times 5$ unitary matrix $\left(U^{(\nu)}_{IJ}\right)$, 
diagonalizing the mass matrix $\left(M^{(\nu)}_{IJ}\right)$ according  to the 
equality  $\,U^{(\nu)\,\dagger}\, M^{(\nu)}\,U^{(\nu)} = {\rm diag}(m_{\nu_1}\,
,\,m_{\nu_2}\,,\,m_{\nu_3}\,,\,m_{\nu_4}\,,\,m_{\nu_5})$~, takes the form

%\vspace{-0.2cm}

%rownanie 13
\begin{equation}
\left({U}^{(\nu)}_{IJ}\right) = \left(\begin{array}{ccccc}
\frac{1}{\sqrt{1+Y^2}} & 0 & 0 & \frac{Y}{\sqrt{1+Y^2}} e^{i\varphi^{(\nu)}} & 
0 \\ 0 & \frac{1}{\sqrt{1+X^2}} & 0 & 0 & \frac{X}{\sqrt{1+X^2}} e^{i
\varphi^{(\nu)}\,'}  \\ 0 & 0 & 1 & 0 & 0 \\ -\frac{Y}{\sqrt{1+Y^2}} e^{-i
\varphi^{(\nu)}} & 0 & 0 & \frac{1}{\sqrt{1+Y^2}} & 0 \\ 0 & -\frac{X}{\sqrt{1+
X^2}} e^{-i\varphi^{(\nu)}\,'} & 0 & 0 & \frac{1}{\sqrt{1+X^2}}
\end{array}\right) \,,
\end{equation}

%\vspace{-0.1cm}

\ni where $ M_{14}^{(\nu)} = |M_{14}^{(\nu)}|\exp  i\varphi^{(\nu)} $, 
$ M_{25}^{(\nu)} = |M_{25}^{(\nu)}|\exp  i\varphi^{(\nu)}\,' $ and

%\vspace{-0.1cm}

%rownanie 14
\begin{eqnarray}
Y &\! =\! & \frac{M_{11}^{(\nu)} - M^{(\nu)}_{44}}{2|M_{14}^{(\nu)}|} + \sqrt{1
+ \left(\frac{M_{11}^{(\nu)} - M^{(\nu)}_{44}}{2|M_{14}^{(\nu)}|}\right)^2} = 
\frac{M_{11}^{(\nu)} - m_{\nu_1}}{|M_{14}^{(\nu)}|} = - \frac{M_{44}^{(\nu)} 
- m_{\nu_4}}{|M_{14}^{(\nu)}|}\;, \nonumber \\ & & \nonumber \\
X &\! =\! & \frac{M_{22}^{(\nu)} - M^{(\nu)}_{55}}{2|M_{25}^{(\nu)}|} + \sqrt{1
+ \left(\frac{M_{22}^{(\nu)} - M^{(\nu)}_{55}}{2|M_{25}^{(\nu)}|}\right)^2} =
\frac{M_{22}^{(\nu)} - m_{\nu_2}}{|M_{25}^{(\nu)}|} = - \frac{M_{55}^{(\nu)} 
- m_{\nu_5}}{|M_{25}^{(\nu)}|}\;. 
\end{eqnarray}

%%\vspace{-0.1cm}

 The neutrino flavor states $\nu_\alpha \equiv \nu_e\,,\,\nu_\mu\,,\,\nu_\tau\,
,\,\nu_s\,,\,\nu_s'$ (of which $\nu_e\,,\,\nu_\mu\,,\,\nu_\tau $, or rather 
their lefthanded parts, stand for the observed weak--interaction neutrino 
states and $\nu_s $, $\nu'_s $ denote their unobserved sterile partners) are 
related to the neutrino mass states $\nu_I \equiv \nu_1\,,\,\nu_2\,,\,\nu_3
\,,\,\nu_4\,,\,\nu_5 $ through a five--dimensional unitary transformation

%\vspace{-0.1cm}

%rownanie 15
\begin{equation}
\nu_\alpha = \sum_J V^*_{J \alpha}\,\nu_J
\end{equation}

%\vspace{-0.1cm}

\ni with $ \left(V^{^*}_{J \alpha}\right)= \left(V_{\alpha J} \right)^\dagger $.
Here,

%\vspace{-0.1cm}

%rownanie 16
\begin{equation}
V_{\alpha J} \equiv \sum_K U^{(\nu)\,*}_{K \alpha}U^{(e)}_{K J} = 
\sum_k U^{(\nu)\,*}_{k\,\alpha} U^{(e)}_{k\,J} + U^{(\nu)\,*}_{4\alpha}
\delta_{4J} + U^{(\nu)\,*}_{5\alpha} \delta_{5J}\,,
\end{equation}

%\vspace{-0.1cm}

\ni where $\left(U^{(e)}_{ij}\right)$ is the charged--lepton diagonalizing 
matrix given in Eq. (9) and 

\vspace{0.1cm}

%rownanie 17
\begin{equation}
U^{(e)}_{i4} = 0 = U^{(e)}_{i5}\;,\;\;U^{(e)}_{4j} = 0 = U^{(e)}_{5j}\;,\;\;
U^{(e)}_{44} = 1 = U^{(e)}_{55}\;.
\end{equation}

\vspace{0.2cm}

\ni The last equations follow from the fact that charged leptons get no 
sterile partners. Thus, from Eq. (16)

\vspace{0.1cm}

%rownanie 18
\begin{equation}
V_{\alpha\,j} = \sum_k U^{(\nu)\,*}_{k\,\alpha}U^{(e)}_{k\,j}\;,\;V_{\alpha4}
= U^{(\nu)\,*}_{4 \alpha}\;,\;V_{\alpha 5} = U^{(\nu)\,*}_{5 \alpha}\; .
\end{equation}

\vspace{0.2cm}

\ni Of course, the $ 5\times 5 $ unitary matrix $\left(V_{\alpha J}\right)$ 
is a five--dimensional lepton counterpart of the familiar CKM  matrix for 
quarks. The charged leptons $ e^-\,,\,\mu^-\,,\,\tau^-$ are here counterparts 
of the up quarks $u\,,\,c\,,t $ (both with diagonalized mass matrix).

 From Eqs. (18), with the use of Eqs. (13) and (9), we can calculate the matrix 
elements $ V_{\alpha J} $ in the lowest (quadratic) perturbative order in 
$\alpha^{(e)}/\mu^{(e)}$. Writing for convenience $\alpha = I = 1,2,3,4,5 $, we
get

\vspace{0.2cm}

%rownanie 19
\begin{eqnarray}
V_{11} & = & \left[1 - \frac{2}{841}\left(\frac{\alpha^{(e)}}{m_\mu}\right)^2
\right]\frac{1}{\sqrt{1+Y^2}}\;, \nonumber \\
V_{22} & = & \left[1 - \frac{2}{841}\left(\frac{\alpha^{(e)}}{m_\mu}\right)^2 
- \frac{96}{841}\left(\frac{\alpha^{(e)}}{m_\tau}\right)^2 \right]
\frac{1}{\sqrt{1+X^2}}\;, \nonumber \\
V_{33} & = & 1 - \frac{96}{841}\left(\frac{\alpha^{(e)}}{m_\tau}\right)^2
\;, \nonumber \\
V_{12} & = & \frac{2}{29} \frac{\alpha^{(e)}}{m_\mu}\,\frac{1}{\sqrt{1+Y^2}}
\,e^{i\varphi^{(e)}}\;\;\;\;\;\;\;,\;\;V_{21} = -\frac{2}{29} \frac{\alpha^{(e)
}}{m_\mu}\,\frac{1}{\sqrt{1+X^2}}\,e^{-i\varphi^{(e)}}\;, \nonumber \\
V_{23} & = & \frac{8\sqrt{3}}{29} \frac{\alpha^{(e)}}{m_\tau} \frac{1}{\sqrt{
1+X^2}}\,e^{i\varphi^{(e)}}\;\;\;\;,\;\;V_{32} =  - \frac{8\sqrt{3}}{29} \frac{
\alpha^{(e)}}{m_\tau} \frac{1}{\sqrt{1+X^2}}\,e^{-i\varphi^{(e)}}\;, \nonumber 
\\ V_{13} & = & 0 \;\;\;\;\;\;\;\;\;\;\;\;\;\;\;\;\;\;\;\;\;\;\;\;\;\;\;\;\;\;
\;\;\;\;\;\;\;\;,\;\;V_{31} = 0 
\end{eqnarray}

%\vspace{-0.3cm}

\ni and

%\vspace{-0.4cm}

%rownanie 20
\begin{eqnarray}
V_{14} & = & -\frac{Y}{\sqrt{1+Y^2}}\,e^{i\varphi^{(\nu)}}\;\;,\;\;V_{41} = 
\left[1 - \frac{2}{841}\left(\frac{\alpha^{(e)}}{m_\mu}\right)^2
\right]\frac{Y}{\sqrt{1+Y^2}}\,e^{-i\varphi^{(\nu)}}\;, \nonumber \\
V_{24} & = & 0\;\;\;\;\;\;\;\;\;\;\;\;\;\;\;\;\;\;\;\;\;\;\;\;\;,\;\;V_{42} =
\frac{2}{29} \frac{\alpha^{(e)}}{m_\mu}\,\frac{Y}{\sqrt{1+Y^2}}\,e^{-i
(\varphi^{(\nu)}-\varphi^{(e)})}\;, \nonumber \\ 
V_{34} & = & 0\;\;\;\;\;\;\;\;\;\;\;\;\;\;\;\;\;\;\;\;\;\;\;\;\;\;,\;\;V_{43} =
0\;\;\;\;\;\;\;\;\;\;\;\;\;\;\;\;\;\;\;\;\;,\;\;V_{44} = \frac{1}{\sqrt{1+Y^2}}
\;, \nonumber \\ 
V_{15} & = & 0\;\;\;\;\;\;\;\;\;\;\;\;\;\;\;\;\;\;\;\;\;\;\;\;\;\;,\;\;V_{51} =
 -\frac{2}{29}\frac{\alpha^{(e)}}{m_\mu}\,\frac{X}{\sqrt{1+X^2}}\,e^{-i
(\varphi^{(\nu)}\,'+\varphi^{(e)})}\;, \nonumber \\ 
V_{25} & = & -\frac{X}{\sqrt{1+X^2}}\,e^{i\varphi^{(\nu)}\,'}\;,\;\;V_{52} = 
\left[1 - \frac{2}{841}\left(\frac{\alpha^{(e)}}{m_\mu}\right)^2-\frac{96}{841}
\left(\frac{\alpha^{(e)}}{m_\tau}\right)^2\right]\frac{X}{\sqrt{1+X^2}}\,e^{-i
\varphi^{(\nu)}\,'}\;, \nonumber \\
V_{35} & = & 0\;\;\;\;\;\;\;\;\;\;\;\;\;\;\;\;\;\;\;\;\;\;\;\;\;\;,\;\;
V_{53} = \frac{8\sqrt{3}}{29} \frac{\alpha^{(e)}}{m_\tau} \frac{X}{
\sqrt{1+X^2}}e^{-i(\varphi^{(\nu)}\,'-\varphi^{(e)})} \;, \nonumber \\ 
V_{45} & = & 0\;\;\;\;\;\;\;\;\;\;\;\;\;\;\;\;\;\;\;\;\;\;\;\;\;\;,\;\;V_{54} =
0\;\;\;\;\;\;\;\;\;\;\;\;\;\;\;\;\;\;\;\;\;,\;\;V_{55} = \frac{1}{\sqrt{1+X^2}}
\;. 
\end{eqnarray}

\vspace{-0.15cm}

\ni In the limit of $\alpha^{(e)} \rightarrow 0$, the only nonzero matrix 
elements $V_{\alpha J}$ are 

\vspace{-0.25cm}

% rownanie 21
\begin{equation}
V_{11} \rightarrow \frac{1}{\sqrt{1+Y^2}} \;,\;\; V_{22} \rightarrow 
\frac{1}{\sqrt{1+X^2}} \;,\;\;V_{33} \rightarrow 1 
\end{equation}

\vspace{-0.15cm}

\ni and

%rownanie 22
\begin{eqnarray}
V_{14} = -\frac{Y}{\sqrt{1+Y^2}} e^{i\varphi^{(\nu)}} \; &\! ,\! & \;\; V_{41} 
\rightarrow -V^*_{14} \;,\;\;V_{44} = \frac{1}{\sqrt{1+Y^2}}\;, \nonumber \\
V_{25} = -\frac{X}{\sqrt{1+X^2}} e^{i\varphi^{(\nu)}\,'} &\! ,\! & \;\; V_{52} 
\rightarrow -V^*_{25}  \;,\;\;V_{55} = \frac{1}{\sqrt{1+X^2}} \;. 
\end{eqnarray}

\vspace{0.4cm}

\ni {\bf 3. Neutrino oscillations}

\vspace{0.4cm}

 Having once found the elements (19) and (20) of the extended lepton CKM 
matrix, we are able to calculate the probabilities of neutrino oscillations 
$\nu_\alpha \rightarrow \nu_\beta $ (in the vacuum), using the familiar 
formula:

\vspace{-0.1cm}

%rownanie 23
\begin{equation}
P(\nu_\alpha \rightarrow \nu_\beta) = |\langle\nu_\beta |\nu_\alpha(t)\rangle
|^2 = \sum_{K\,L}V_{L\,\beta}V^*_{L\,\alpha}V^*_{K\,\beta}V_{K\,\alpha} 
\exp\left(i\frac{m^2_{\nu_L}-m^2_{\nu_K}}{2|\vec{p}|}\,t\right)\;,
\end{equation}

\vspace{-0.1cm}

\ni where $\nu_\alpha(0) = \nu_\alpha $, $\langle\nu_\beta| =\langle 0 |
\nu_\beta $ and $\langle\nu_\beta |\nu_\alpha \rangle = \delta_{\beta\,\alpha}
$. Here, as usual, $t/|\vec{p}| = L/E\,\;( c = 1 = \hbar)$, what is equal to 
$ 4\times 1.2663 L/E $ if $m^2_{\nu_L} - m^2_{\nu_K}\;,\; L $ and $ E $ are
measured in eV$^2$, m and MeV, respectively. Of course, $ L $ is the source--%
detector distance (the baseline). In the following, it will be convenient to 
denote 

\vspace{-0.2cm}

%rownanie 24
\begin{equation}
x_{LK} = 1.2663 \frac{(m^2_{\nu_L} - m^2_{\nu_K}) L}{E}  
\end{equation} 

\vspace{-0.1cm}
 
\ni and use the identity $ \cos 2x_{LK} = 1 - 2\sin^2 x_{LK}$.

 From Eqs. (23), (19) and (20) we derive by explicit calculations the following
neutrino--oscillation formulae valid in the lowest (quadratic) perturbative 
order in $\alpha^{(e)}/\mu^{(e)}$:

\vspace{-0.2cm}

%rownanie 25
\begin{eqnarray}
\lefteqn{P\left(\nu_e \rightarrow \nu_\mu  \right) = \frac{16}{841} 
\left(\frac{\alpha^{(e)}}{m_\mu}\right)^2 } \nonumber \\ 
& & \times \left[\frac{1}{(1+X^2)(1+Y^2)}\left(\sin^2 x_{21}+X^2 \sin^2 x_{51}
+Y^2 \sin^2 x_{42}+ X^2 Y^2 \sin^2 x_{54}\right)\right. \nonumber \\ 
& & \;\;\;-\left. \frac{X^2}{(1+X^2)^2}\sin^2 x_{52} - \frac{Y^2}{(1+Y^2)^2}
\sin^2 x_{41} \right]\;,\nonumber \\
\lefteqn{P\left(\nu_e \rightarrow \nu_\tau \right) = 0 \;, } \nonumber \\
\lefteqn{P\left(\nu_\mu \rightarrow \nu_\tau \right) = \frac{768}{841}\!\! 
\left(\frac{\alpha^{(e)}}{m_\tau}\right)^2\!\left[ \frac{1}{1\!+\!X^2}\! 
\left( \sin^2\! x_{32}\!\! +\!\! X^2 \sin^2 x_{53}\right)\!\! -\!\!
\frac{X^2}{(1\!+\!X^2)^2} \sin^2\! x_{52}\right] \;,} \nonumber \\
\lefteqn{P\left(\nu_e \rightarrow \nu_s\right) = 4\left[1 - \frac{4}{841}
\left( \frac{\alpha^{(e)}}{m_\mu}\right)^2 \right] \frac{Y^2}{(1+Y^2)^2}
\sin^2 x_{41} \;,} \nonumber \\
\lefteqn{P\left(\nu_\mu \rightarrow \nu_s\right) = \frac{16}{841}\left(
\frac{\alpha^{(e)}}{m_\mu}\right)^2 \frac{Y^2}{(1+Y^2)^2} \sin^2 x_{41} \;,}
\nonumber \\
\lefteqn{P\left(\nu_e \rightarrow \nu'_s\right) = \frac{16}{841} \left(
\frac{\alpha^{(e)}}{m_\mu}\right)^2 \frac{X^2}{(1+X^2)^2} \sin^2 x_{52}\;,} 
\nonumber \\ 
\lefteqn{P\left(\nu_\mu \rightarrow \nu'_s\right) = 4\left[1\!\! -\!\! 
\frac{4}{841}\left(\frac{\alpha^{(e)}}{m_\mu}\right)^2\!\! -\!\!\frac{192}{841}
\left(\frac{\alpha^{(e)}}{m_\tau}\right)^2 \right] \frac{X^2}{(1\!+\!X^2)^2} 
\sin^2\! x_{52} \;.}
\end{eqnarray}

\vspace{-0.2cm}

\ni In the limit of $\alpha^{(e)}\rightarrow 0$, the only nonzero neutrino--%
oscillation probabilities are

%rownanie 26
\begin{eqnarray}
P\left(\nu_e \rightarrow \nu_s\right) & \rightarrow & 4 \frac{Y^2}{(1+Y^2)^2} 
\sin^2 x_{41} \;, \nonumber \\
P\left(\nu_\mu \rightarrow \nu'_s\right) & \rightarrow & 4\frac{X^2}{(1+X^2)^2}
\sin^2 x_{52} \;.
\end{eqnarray}

 The formulae (25) for the disappearance modes of $\nu_e $ and $\nu_\mu $
imply the following survival probabilities for $\nu_e $ and $\nu_\mu $:

\vspace{-0.2cm}

%rownanie 27 
\begin{eqnarray}
\lefteqn{ P\left(\nu_e \rightarrow \nu_e\right) = 1\! - \!P\left(\nu_e 
\rightarrow \nu_\mu\right)\! - \!P\left(\nu_e \rightarrow \nu_\tau\right)\! - 
\!P\left(\nu_e \rightarrow \nu_s \right)\! - \!P\left(\nu_e \rightarrow \nu'_s
\right) } \nonumber \\
& & = 1 - 4\left[1 - \frac{8}{841}\left(\frac{\alpha^{(e)}}{m_\mu}\right)^2 
\right] \frac{Y^2}{(1+Y^2)^2} \sin^2 x_{41}  \nonumber \\
& &\;\;\; - \frac{16}{841}\left(\frac{\alpha^{(e)}}{m_\mu}\right)^2\! 
\frac{1}{(1+X^2)(1+Y^2)} \left(\sin^2 \!x_{21}\!\! +\!\! X^2 \sin^2 \!x_{51}\!
\! +\!\! Y^2 \sin^2 \!x_{42}\!\! +\!\! X^2 Y^2 \sin^2 \!x_{54} \right) 
\nonumber \\ 
\end{eqnarray}

\vspace{-0.2cm}

\ni and

%rownanie 28 
\begin{eqnarray}
\lefteqn{P\left(\nu_\mu \rightarrow \nu_\mu\right) = 1 - P\left(\nu_\mu \rightarrow
\nu_e\right) - P\left(\nu_\mu \rightarrow \nu_\tau\right) - P\left(\nu_\mu 
\rightarrow \nu_s\right) - P\left(\nu_\mu \rightarrow \nu'_s\right)} \nonumber 
\\& & = 1 - 4\left[1 - \frac{8}{841}\left(\frac{\alpha^{(e)}}{m_\mu}\right)^2 
- \frac{384}{841}\left(\frac{\alpha^{(e)}}{m_\tau}\right)^2\right] 
\frac{X^2}{(1+X^2)^2} \sin^2 x_{52}  \nonumber \\
& & \;\;\;- \frac{16}{841}\left(\frac{\alpha^{(e)}}{m_\mu}\right)^2\! \frac{1}{
(1+X^2)(1+Y^2)}\left(\sin^2\! x_{21}\!\! +\!\! X^2\sin^2\! x_{51}\!\! +\!\! Y^2
\sin^2\! x_{42} +X^2 Y^2 \sin^2\! x_{54} \right) . \nonumber \\ 
\end{eqnarray}

\ni In the limit of $\alpha^{(e)}\rightarrow 0$, we obtain

%rownanie 29
\begin{equation}
P\left(\nu_e \rightarrow \nu_e\right) \rightarrow 1 - 4 \frac{Y^2}{(1+Y^2)^2} 
\sin^2 x_{41}
\end{equation}

\ni and

%rownanie 30
\begin{equation}
P\left(\nu_\mu \rightarrow \nu_\mu\right) \rightarrow 1 - 4 \frac{X^2}{(1
+X^2)^2} \sin^2 x_{52}\;.
\end{equation}

 The last two formulae are to be compared with solar--neutrino and atmospheric%
--neutrino experiments, respectively.

%\vfill\eject

\vspace{0.4cm}

\ni {\bf 4. Atmospheric and solar neutrinos}

\vspace{0.4cm}

In the case of atmospheric neutrinos, we compare our formula (30) with Eq. (1).
Then, for instance, 

%rownanie 31
\begin{equation} 
\frac{4 X^2}{(1+X^2)^2} \sim 0.9
\end{equation}

\ni (more generally: $\sim 0.82 $ to 1) and

%rownanie 32
\begin{equation}
m^2_{\nu_5} - m^2_{\nu_2} \sim 5 \times 10^{-3}\;\;{\rm eV}^2
\end{equation}

\ni (more generally: $\sim (0.5\;\;{\rm to}\;\;6)\times 10^{-3}\;\;
{\rm eV}^2$).

 From the input (31) we get

%rownanie 33
\begin{equation}
X \sim 0.721
\end{equation}

\ni and, through the second Eq. (14),

%rownanie 34
\begin{equation}
\frac{M^{(\nu)}_{55} - M^{(\nu)}_{22}}{2|M^{(\nu)}_{25}|} = \frac{1 - X^2}{2X}
\sim \frac{1}{3}
\end{equation}

\ni or

%rownanie 35
\begin{equation}
|M^{(\nu)}_{25}| = \frac{X}{1 - X^2}\left( M^{(\nu)}_{55} - M^{(\nu)}_{22} 
\right) \sim  \frac{3}{2} \left( M^{(\nu)}_{55} - M^{(\nu)}_{22} \right)\;.
\end{equation}

\ni On the other hand, the third mass formula (12) and the input (32) give

%rownanie 36
\begin{equation}
\left( M^{(\nu)}_{22} + M^{(\nu)}_{55} \right) \sqrt{ \left(M^{(\nu)}_{22} - 
M^{(\nu)}_{55} \right)^2 + 4|M^{(\nu)}_{25}|^2 } =
m^2_{\nu_5} - m^2_{\nu_2} \sim 5 \times 10^{-3}\;\;{\rm eV}^2
\end{equation}

\ni or, with the use of Eqs. (34) and (33),

%rownanie 37
\begin{equation}
M^{(\nu)\,2}_{55}-M^{(\nu)\,2}_{22}= \frac{1 -X^2}{1 + X^2}\left(m^2_{\nu_5} - 
m^2_{\nu_2}\right) \sim 1.58 \times 10^{-3}\;\;{\rm eV}^2 \;.
\end{equation}

\ni With the formulae $ M^{(\nu)}_{22} \simeq 320 \mu^{(\nu)}/261 $ and 
$ M^{(\nu)}_{55} \sim 48 \mu^{(\nu)}/7 $ we have $ M^{(\nu)\,2}_{55} - 
M^{(\nu)\,2}_{22} \sim 45.5 \mu^{(\nu)\,2} $. Hence, Eq. (37) leads to

%rownanie 38
\begin{equation}
\mu^{(\nu)} \sim 5.90 \times 10^{-3}\;\;{\rm eV}.
\end{equation}

\ni Then,

\vspace{-0.1cm}

%rownanie 39
\begin{equation}
M^{(\nu)}_{22} \sim 7.25 \times 10^{-3}\;\;{\rm eV}\;\;,\;\;
M^{(\nu)}_{55} \sim 4.04 \times 10^{-2}\;\;{\rm eV}
\end{equation}

\vspace{-0.1cm}

\ni and so, from Eq. (35)

\vspace{-0.1cm}

%rownanie 40
\begin{equation}
|M^{(\nu)}_{25}| \sim 4.97 \times 10^{-2}\;\;{\rm eV}.
\end{equation}

\vspace{-0.1cm}

\ni Finally, with the values (39) and (40) the third mass formula (12) gives

\vspace{-0.1cm}

%rownanie 41
\begin{equation}
m_{\nu_2,\nu_5} \sim \left\{ \begin{array}{r} -2.86 \times 10^{-2}\;\;{\rm eV}
\\ 7.62 \times 10^{-2}\;\;{\rm eV} \end{array} \right.\;.
\end{equation}

\vspace{-0.1cm}

 In this way, all parameters appearing in our model of neutrino "texture", 
needed to explain the observed deficit of atmospheric $\nu_\mu$'s in terms of
neutrino oscillations $\nu_\mu \rightarrow \nu'_s $, are determined.

 In the case of solar neutrinos, we compare our formula (29) with the survival
probability for $\nu_e $, usually analized experimentally in two--flavor form

\vspace{-0.1cm}

%rownanie 42
\begin{equation}
P\left(\nu_e \rightarrow \nu_e\right) = 1 - \sin^2 2\theta_{\rm sol}
\sin^2 \left(1.27 \Delta m^2_{\rm sol}\, L/E \right)\;.
\end{equation}

\vspace{-0.1cm}

\ni Taking into account the so--called vacuum fit [7] ({\it i.e.}, one that is 
not enhanced by the resonant MSW mechanism [8] in the Sun matter), we have the 
parameters

\vspace{-0.1cm}

%rownanie 43
\begin{equation}
\sin^2 2\theta_{\rm sol} \sim 0.65\;\;{\rm to}\;\;1\;\;,\;\;\Delta m^2_{\rm 
sol} \sim (5\;\;{\rm to}\;\;8) \times 10^{-11}\;{\rm eV}^2 \;,
\end{equation}

\vspace{-0.1cm}

\ni what shows a large mixing and a very small difference of masses squared. 
Then, for instance,

\vspace{-0.1cm}

%rownanie 44
\begin{equation}
\frac{4 Y^2}{(1+Y^2)^2}  \sim 0.8
\end{equation}

\vspace{-0.1cm}

\ni (more generally: $\sim 0.65 $ to 1) and

\vspace{-0.1cm}

%rownanie 45
\begin{equation}
m^2_{\nu_4} - m^2_{\nu_1} \sim 7 \times 10^{-11}\;{\rm eV}^2 
\end{equation}

\vspace{-0.1cm}

\ni (more generally: $\sim (5$ to 8) $\times 10^{-11}\;{\rm eV}^2$). 

 From the input (44) we obtain 

\vspace{-0.1cm}

%rownanie 46
\begin{equation}
Y \simeq 0.618
\end{equation}

\vspace{-0.1cm}

\ni and, due to the first Eq. (14),

\vspace{-0.1cm}

%rownanie 47
\begin{equation}
\frac{M^{(\nu)}_{44} - M^{(\nu)}_{11}}{2|M^{(\nu)}_{14}|} = \frac{1 - Y^2}{2Y} 
\sim \frac{1}{2}
\end{equation}

\vspace{-0.1cm}

\ni or 

\vspace{-0.1cm}

%rownanie 48
\begin{equation}
|M^{(\nu)}_{14}| = \frac{Y}{1 - Y^2}\left(M^{(\nu)}_{44} - M^{(\nu)}_{11}
\right) \sim M^{(\nu)}_{44} - M^{(\nu)}_{11}\;.
\end{equation}

\vspace{-0.1cm}

\ni On the other hand, the first mass formula (12) and the input (45) lead to

\vspace{-0.1cm}

%rownanie 49
\begin{equation}
\left( M^{(\nu)}_{11} + M^{(\nu)}_{44} \right) \sqrt{ \left(M^{(\nu)}_{11} - 
M^{(\nu)}_{44} \right)^2 + 4|M^{(\nu)}_{14}|^2 } =
m^2_{\nu_4} - m^2_{\nu_1} \sim 7 \times 10^{-11}\;\;{\rm eV}^2
\end{equation}

\vspace{-0.1cm}

\ni or, through Eqs. (47) and (46), to

\vspace{-0.1cm}

%rownanie 50
\begin{equation}
M^{(\nu)\,2}_{44} -M^{(\nu)\,2}_{11} = \frac{1 - Y^2}{1 + Y^2}\left(m^2_{\nu_4}
- m^2_{\nu_1}\right) \sim 3.13 \times 10^{-11}\;{\rm eV}^2\;.
\end{equation}

\vspace{-0.1cm}

\ni With the formulae $ M^{(\nu)}_{11} = \mu^{(\nu)}\varepsilon^{(\nu)\,2}/29 
$ and $ M^{(\nu)}_{44} \sim \mu^{(\nu)}\varepsilon^{(\nu)\,2}/7 $ we get 
$ M^{(\nu)\,2}_{44} - M^{(\nu)\,2}_{11} \sim 0.0192 \mu^{(\nu)\,2}\varepsilon^{
(\nu)\,4} $. Hence, Eqs. (50) and (38) give

\vspace{-0.1cm}

%rownanie 51
\begin{equation}
\varepsilon^{(\nu)\,2} \sim 6.85 \times 10^{-3}\;.
\end{equation}

\vspace{-0.2cm}

\ni Then,

\vspace{-0.2cm}

%rownanie 52
\begin{equation}
M^{(\nu)}_{11} \sim 1.39 \times 10^{-6}\;{\rm eV}\;\;,\;\;M^{(\nu)}_{44} \sim 
5.77 \times 10^{-6}\;{\rm eV}
\end{equation}

\vspace{-0.1cm}

\ni and thus, from Eq. (48)

\vspace{-0.1cm}

%rownanie 53
\begin{equation}
|M^{(\nu)}_{14}| \sim 4.38 \times 10^{-6}\;{\rm eV}\;.
\end{equation}

\vspace{-0.1cm}

\ni Eventually, with the values (52) and (53) the first mass formula (12) 
implies

\vspace{-0.1cm}

%rownanie 54
\begin{equation}
m_{\nu_1,\nu_4} \sim \left\{ \begin{array}{r} -1.32 \times 10^{-6}\;\;{\rm eV}
\\ 8.48 \times 10^{-6}\;\;{\rm eV} \end{array} \right.\;.
\end{equation}

\vspace{-0.1cm}

 In such a way, all parameters contained in our model of neutrino "texture",
needed to describe the observed deficit of solar $\nu_e$'s in terms of neutrino
oscillations $\nu_e \rightarrow \nu_s $ in the vacuum, are determined.

Our last item is concerned with the LSND accelerator experiment that reported 
the detection of $\bar{\nu}_\mu \rightarrow \bar{\nu}_e $ and $\nu_\mu 
\rightarrow \nu_e $ oscillations by observing $\bar{\nu}_e $'s and $\nu_e $'s 
in a beam of $\bar{\nu}_\mu $'s and $\nu_\mu $'s produced in $\pi^-$ and 
$\pi^+$ decays, respectively [9]. The observed excess of $\bar{\nu}_e $'s and 
$\nu_e$'s, analized in terms of two--flavor neutrino--oscillation formula, 
implies a considerable amplitude $\sin^2 2\theta_{\rm LSND}$, too large to be 
explained by our formula (25) for $P(\nu_\mu \rightarrow \nu_e) = P(\nu_e 
\rightarrow \nu_\mu)$, where the amplitude at $\sin^2 x_{21} $, 

\vspace{-0.1cm}

%rownanie 55
\begin{equation}
\frac{16}{841}\left(\frac{\alpha^{(e)}}{m_\mu}\right)^2\frac{1}{(1 + X^2)
(1 + Y^2)}\;,
\end{equation}

\vspace{-0.15cm}

\ni is small:

\vspace{-0.25cm}

%rownanie 56
\begin{equation}
0 \leq\frac{16}{841}\left(\frac{\alpha^{(e)}}{m_\mu}\right)^2 \leq 
6.2 \times 10^{-4} \;,
\end{equation}

\vspace{-0.1cm}

\ni as it follows from Eq. (8). Here, the central value is

\vspace{-0.2cm}

%rownanie 57
\begin{equation}
\frac{16}{841}\left(\frac{\alpha^{(e)}}{m_\mu}\right)^2 = 2.5\times 10^{-4}
\;.
\end{equation}

\vspace{0.1cm}

 I would like to thank Jan Kr\'{o}likowski for several helpful discussions.

%\vfill\eject

\vspace{0.6cm}

{\centerline{\bf Appendix: Unified "texture dynamics"}}

\vspace{0.4cm}

\appendix\setcounter{equation}{0}

 In this Appendix the idea of a model of fermion "texture" that we develop 
since some time [4,5] is outlined. In particular, the existence of two sterile 
neutrinos $\nu_s $ and $\nu'_s $ turns out to follow naturally.

 Let us introduce the following $ 3\times 3 $ matrices in the space of three 
fermion families: 

\vspace{0.1cm}

%(A.1)
$$
\widehat{a} = \left(\begin{array}{ccc} 0 & 1 & 0 \\ 0 & 0 & \sqrt{2} \\ 0 & 0 
& 0 \end{array} \right)\;\;,\;\;\widehat{a}^\dagger = \left(\begin{array}{ccc} 
0 & 0 & 0 \\ 1 & 0 & 0 \\ 0 & \sqrt{2} & 0 \end{array} \right)\;\;.
\eqno({\rm A}.1)
$$

\vspace{0.25cm}

\ni With the matrix

\vspace{0.1cm}

%(A.2)
$$
\widehat{n} = \widehat{a}^\dagger\widehat{a} = \left(\begin{array}{ccc} 
0 & 0 & 0 \\ 0 & 1 & 0 \\ 0 & 0 & 2 \end{array} \right)\;\;, \eqno({\rm A}.2)
$$

\vspace{0.25cm}

\ni they satisfy the commutation relations

\vspace{0.1cm}

%(A.3)
$$
[\widehat{a}\,,\,\widehat{n}] = \widehat{a}\;,\;[\widehat{a}^\dagger\,,
\,\widehat{n}] = -\widehat{a}^\dagger \eqno({\rm A}.3)
$$

\vspace{0.25cm}

\ni characteristic for annihilation and creation matrices, while $\widehat{n}$ 
plays the role of an occup\-ation--\-number matrix. However, in addition, they 
obey the "truncation" identities

\vspace{0.1cm}

%(A.4)
$$
\widehat{a}^3 = 0\,,\, \widehat{a}^{\dagger\,3} = 0\,. \eqno({\rm A}.4)
$$

\vspace{0.25cm}

\ni Note that due to Eqs. (A.4) the bosonic canonical commutation relation
$[\widehat{a}\,,\,\widehat{a}^\dagger] = \widehat{1}$ does not hold, being 
replaced by the relation $[\widehat{a}\,,\,\widehat{a^\dagger}] = $ diag~$(1\,,
\,1\,,\,-2)$.

 In consequence of Eqs. (A.1), (A.2) and (A.3), we get $\widehat{n}|n\rangle = 
n|n\rangle $ as well as $\widehat{a}|n\rangle = \sqrt{n}|n-1\rangle $
and $\widehat{a}^\dagger|n\rangle = \sqrt{n+1}|n+1\rangle\;\;(n = 0,1,2) $,
however, $\widehat{a}^\dagger|2\rangle = 0 $ ({\it i.e.}, $|3\rangle = 0 $) in
addition to $\widehat{a}^\dagger|0\rangle = 0 $ ({\it i.e.}, $|-1\rangle = 0$).
Evidently, $n = 0,1,2 $ may play the role of a vector index in our three--%
dimensional matrix calculus.

 It is natural to expect that the Gell--Mann matrices (generating the horizon%
tal SU(3) algebra) can be built up from $\widehat{a}$ and $\widehat{a}^\dagger
$. In fact,

\vfill\eject

\vspace{-0.15cm}

%(A.5)
\begin{eqnarray*}
\widehat{\lambda}_1 & = & \left(\begin{array}{rrr} 0 & 1 & 0 \\ 1 & 0 & 0 \\
0 & 0 & 0 \end{array}\right) = \frac{1}{2}\left(\widehat{a}^2\widehat{a}^{
\dagger} + \widehat{a}\widehat{a}^{\dagger\,2} \right)\;,\\
\widehat{\lambda}_2 & = & \left(\begin{array}{rrr} 0 & -i & 0 \\ i & 0 & 0 \\
0 & 0 & 0 \end{array}\right) = \frac{1}{2i}\left(\widehat{a}^2\widehat{a}^{
\dagger} - \widehat{a}\widehat{a}^{\dagger\,2} \right)\;,\\
\widehat{\lambda}_3 & = & \left(\begin{array}{rrr} 1 & 0 & 0 \\ 0 & -1 & 0  \\
0 & 0 & 0 \end{array}\right) = \frac{1}{2}\left(\widehat{a}^2\widehat{a}^{
\dagger\,2} - \widehat{a}\widehat{a}^{\dagger\,2}\widehat{a} \right)\;,\\
\widehat{\lambda}_4 & = & \left(\begin{array}{rrr} 0 & 0 & 1 \\ 0 & 0 & 0 \\
1 & 0 & 0 \end{array}\right) = \frac{1}{\sqrt{2}}\left(\widehat{a}^2 + 
\widehat{a}^{\dagger\,2} \right)\;,\\
\widehat{\lambda}_5 & = & \left(\begin{array}{rrr} 0 & 0 & -i \\ 0 & 0 & 0 \\
i & 0 & 0 \end{array}\right) = \frac{1}{i\,\sqrt{2}}\left(\widehat{a}^2 - 
\widehat{a}^{\dagger\,2} \right)\;,\\
\widehat{\lambda}_6 & = & \left(\begin{array}{rrr} 0 & 0 & 0 \\ 0 & 0 & 1 \\
0 & 1 & 0 \end{array}\right) = \frac{1}{\sqrt{2}}\left(\widehat{a}^\dagger
\widehat{a}^2 + \widehat{a}^{\dagger \,2}\widehat{a}\right)\;,\\
\widehat{\lambda}_7 & = & \left(\begin{array}{rrr} 0 & 0 & 0 \\ 0 & 0 & -i \\
0 & i & 0 \end{array}\right) = \frac{1}{i\sqrt{2}}\left(\widehat{a}^\dagger
\widehat{a}^2 - \widehat{a}^{\dagger \,2}\widehat{a} \right)\;,\\
\widehat{\lambda}_8 & = & \frac{1}{\sqrt{3}}\,\left(\begin{array}{rrr} 1 & 0 & 
0 \\ 0 & 1 & 0 \\ 0 & 0 & -2 \end{array}\right) = \frac{1}{\sqrt{3}}\left(
\widehat{a}\widehat{a}^\dagger - \widehat{a}^\dagger\widehat{a} \right)\,,
\\
\widehat{1} & = & \left(\begin{array}{rrr} 1 & 0 & 0 \\ 0 & 1 & 0 \\ 0 & 0 & 1
\end{array}\right) = \frac{1}{2}\left(\widehat{a}^2\widehat{a}^{\dagger\,2} 
+ \widehat{a}\widehat{a}^{\dagger\,2}\widehat{a} + \widehat{a}^{\dagger\,2}
\widehat{a}^2\right)\;.
\end{eqnarray*}

\vspace{-1.92cm}

\begin{flushright}
(A.5)
\end{flushright}

\vspace{0.5cm}

\ni Inversely, $\widehat{a} = (\widehat{\lambda}_1 + i\widehat{\lambda}_2)/2 +
\sqrt{2}(\widehat{\lambda}_6+i\widehat{\lambda}_7)/2$ and $\widehat{a}^{\dagger
} = (\widehat{\lambda}_1 - i\widehat{\lambda}_2)/2+ \sqrt{2}(\widehat{\lambda}
_6 - i\widehat{\lambda}_7)/2$. A message we get from these relationships is 
that a horizontal field formalism, always simple (linear) in terms of $ 
\widehat{\lambda}_A\;\;(A = 1,2,\ldots,8)$ and $\widehat{1}$, is generally not 
simple in terms of $\widehat{a}$ and $\widehat{a}^{\dagger}$. In particular, a 
nontrivial SU(3)--symmetric horizontal formalism is not simple in $\widehat{a}$
and $\widehat{a}^{\dagger}$. Inversely, a nontrivial horizontal field formal\-%
ism, if simple (linear and/or quadratic and/or cubic) in terms of $\widehat{a}$
and $\widehat{a}^{\dagger}$, cannot be SU(3)--symmetric.

 Now, let us consider the following ansatz [5]:

\vspace{0.15cm}

%(A.6)
$$
\widehat{M}^{(f)} = \widehat{\rho}^{\,1/2}\widehat{h}^{(f)}\widehat{\rho}^{\,
1/2} \;\;\;(f = \nu\,,\,e\,,\,u\,,\,d)\;, \eqno({\rm A}.6)
$$

\vspace{0.15cm}

\ni where

\vspace{0.15cm}

%(A.7)
$$ 
\widehat{\rho}^{\,1/2} = \frac{1}{\sqrt{29}}
\left(\begin{array}{rrr} 1 & 0 & 0 \\ 0 & \sqrt{4} & 0 
\\ 0 & 0 & \sqrt{24} \end{array}\right)\;\;,\;\; {\rm Tr}\widehat \rho = 1
  \eqno({\rm A}.7)
$$

\vspace{0.15cm}

\ni and 

\vspace{0.15cm}

%(A.8)
\begin{eqnarray*}
\widehat{h}^{(f)} & = & \mu^{(f)}\left[\left(1 + 2\widehat{n}\right)^2 
+ \left(\varepsilon^{(f)\,2} -1 \right)\left(1 + 2\widehat{n}\right)^{-2} + 
\widehat{C}^{(f)}\right] \\
& + & \left(\alpha^{(f)}\widehat{1} - \beta^{(f)}
\widehat{n}\right)\widehat{a}e^{i\varphi^{(f)}} + \widehat{a}^\dagger\left(
\alpha^{(f)}\widehat{1} - \beta^{(f)}\widehat{n}\right)e^{-i\varphi^{(f)}}
\end{eqnarray*}    

\vspace{-1.55cm}

\begin{flushright}
(A.8)
\end{flushright}

\vspace{0.15cm}

\ni with $\widehat{n} = \widehat{a}^\dagger\widehat{a}$ and

\vspace{0.15cm}

%(A.9)
$$
\widehat{1} + 2 \widehat{n} = \widehat{N} = \left(\begin{array}{rrr} 
1 & 0 & 0 \\ 0 & 3 & 0 \\ 0 & 0 & 5 \end{array}\right)\;\,,\,\;
\widehat{C}^{(f)} =\left(\begin{array}{lll} 0 & 0 & 0 \\ 0 & 0 & 0 \\ 0 & 0 & 
C^{(f)} \end{array}\right)\;.   \eqno({\rm A}.9)
$$

\vspace{0.15cm}

\ni It is the matter of an easy calculation to show that the matrices (A.6)
get explicitly the form [5]:

\vspace{0.15cm}

%rownanie (A.10)
$$
\widehat{M}^{(f)} = \frac{1}{29} \left(\begin{array}{ccc} 
\mu^{(f)}\varepsilon^{(f)\,2} & 2\alpha^{(f)} e^{i\varphi^{(f)}} & 0 \\ & & 
\\ 2\alpha^{(f)} e^{-i\varphi^{(f)}} & 4\mu^{(f)}(80 + \varepsilon^{(f)\,2})/9 
& 8\sqrt{3}(\alpha^{(f)}\! - \!\beta^{(f)}) e^{i\varphi^{(f)}} \\ & & \\ 0 & 8
\sqrt{3}(\alpha^{(f)}\! - \!\beta^{(f)}) e^{-i\varphi^{(f)}} & 24\mu^{(f)}(624
\! +\! 25C^{(f)}+ \varepsilon^{(f)\,2})/25 \end{array}\right)\,.
\eqno({\rm A}.10)
$$

\vspace{0.15cm}

\ni In this paper we write also $\widehat{M}^{(f)} = \left(M^{(f)}_{ij}
\right)\;\;(i,j = 1,2,3)$.

 In a more detailed construction following from our idea about the origin of 
three fermion families [4], each eigenvalue $ N = 1\,,\,3\,,\,5 $ of the matrix
$\widehat{N}$ corresponds (for any $f = \nu\,,\,e\,,\,u\,,\,d$) to a wave func%
tion carrying $ N = 1\,,\,3\,,\,5 $ Dirac bispinor indices: $\alpha_1,\alpha_2,
\ldots,\alpha_N $ of which one, say $\alpha_1 $, is coupled to the external 
Standard--Model gauge fields, while the remaining $ N-1 = 0\,,\,2\,,\,4\,:\;\,
\alpha_2\,,\ldots,\;\alpha_N $ (that are not coupled to these fields) are fully
antisymmetric under permutations. So, the latter obey Fermi statistics along 
with the Pauli principle implying that really $ N-1 \leq 4$, because each 
$\alpha_i = 1,2,3,4 $. Then, the three wave functions corresponding to $ N = 
1\,,\,3\,,\,5 $ can be reduced to three other wave functions carrying only one 
Dirac bispinor index $\alpha_1$ (and so, spin 1/2),

%rownanie (A.11)
\begin{eqnarray*}
\psi^{(f)}_{1\,\alpha_1} & \equiv & \psi^{(f)}_{\alpha_1}\;, \\
\psi^{(f)}_{3\,\alpha_1} & \equiv & \frac{1}{4}\left(C^{-1}\gamma^5 \right)_{
\alpha_2\alpha_3} \psi^{(f)}_{\alpha_1\alpha_2\alpha_3}  = \psi^{(f)}_{\alpha_1
\,12} =  \psi^{(f)}_{\alpha_1\,34}\;, \\
\psi^{(f)}_{5\,\alpha_1} & \equiv & \frac{1}{24}\varepsilon_{\alpha_2\alpha_3
\alpha_4\alpha_5}\psi^{(f)}_{\alpha_1\alpha_2\alpha_3\alpha_4\alpha_5} =
\psi^{(f)}_{\alpha_1\,1234} \;,
\end{eqnarray*} 

\vspace{-1.53cm}

\begin{flushright}
(A.11)
\end{flushright}

\ni and appearing (up to the sign) with the multiplicities 1, 4 and 24, 
respectively. In this argument, for $ N = 3 $ the requirement of relativistic 
covariance of the wave function (and the related probability current) is 
applied explicitly [4]. The weighting matrix $\widehat{\rho}^{\,1/2}$ as given 
in Eq. (A.7) gets as its elements the square roots of these multiplicities, 
normalized in such a way that Tr$\,\widehat{\rho} = 1 $.

 In Eqs. (A.11), the indices $\alpha_i $ ($i = 1,2,\ldots,N $) are of Jacobi 
type: $\alpha_1 $ is a "centre--of--mass" Dirac bispinor index, while $\alpha_2
\,,\ldots,\;\alpha_N $ are "relative" Dirac bispinor indices. In fact, $
\alpha_i $ ($i = 1,2,\ldots,N$) are defined by chiral representations of $
\Gamma^\mu_i$ matrices ($i = 1,2,\ldots,N$) being the (properly normalized) 
Jacobi combinations of some individual $\gamma^\mu_i$ matrices ($i = 1,2,\ldots,
N$), where, in particular, $\Gamma^\mu_1 \equiv (1/\sqrt{N}) \sum_{i=1}^N 
\gamma^\mu_i$ [4]. For them $\left\{\Gamma^\mu_i,\Gamma^\nu_j\right\} = 2\delta
_{ij} g^{\mu \nu} $ ($i,j = 1,2,\ldots,N$), in consequence of the anticom\-%
mutation relations $\left\{\gamma^\mu_i,\gamma^\nu_j\right\} = 2\delta_{ij} 
g^{\mu \nu} $ valid for any $\gamma^\mu_i $ and $\gamma^\nu_j $. Then, the 
Dirac--type equations $\left\{\Gamma_1\cdot\left[ p - g A(x)\right] - M\right\}
\psi (x) = 0 $ ($N = 1,2,3,\ldots $) [4], independent of $\Gamma^\mu_2\,,
\ldots,\;\Gamma^\mu_N $, hold for the fundamental--particle wave functions $
\psi (x) = \left(\psi_{\alpha_1\alpha_2\ldots\alpha_N}(x) \right)$, where $ N 
= 1,3,5 $ in the case of fermion wave functions (A.11). Here, $ g\Gamma_1 \cdot
A(x) $ symbolizes the Standard--Model coupling.

 Note that all four matrices $\widehat{M}^{(f)}\;\;(f = \nu\,,\,e\,,\,u\,,\,d)$
defined by Eqs. (A.6) --- (A.9) and (A.1) have a common structure, differing 
from each other only by the values of their parameters $\mu^{(f)}$, $
\varepsilon^{(f)\,2}$, $\alpha^{(f)}$, $\beta^{(f)}$, $ C^{(f)}$ and $\varphi^{
(f)}$. We proposed the fermion mass matrices to be of this unified form [5]. 
Then, Eqs. (A.6) and (A.8) define a quantum--mechanical model for the "texture" 
of fermion mass matrices $\widehat{M}^{(f)}\;\;(f = \nu\,,\,e\,,\,u\,,\,d)$. 
Such an approach may be called "texture dynamics".

 The fermion mass matrix $\widehat{M}^{(f)}$, containing the kernel $\widehat{h
}^{(f)}$ given in Eq. (A.8), consists of a diagonal part proportional to  $
\mu^{(f)}$, and of an off--diagonal part involving linearly $\alpha^{(f)}$ and 
$\beta^{(f)}$. The off--diagonal part of $\widehat{h}^{(f)}$ describes the 
mixing of three eigenvalues

\vspace{-0.1cm}

%(A.12)
$$
\mu^{(f)}\left[N^2 + \left(\varepsilon^{(f)\,2} - 1\right)N^{-2} + \delta_{N\,
5} C^{(f)}\right]\;\;(N = 1,3,5)  \eqno({\rm A}.12)
$$

\vspace{-0.1cm}

\ni of its diagonal part. Beside the term  $\mu^{(f)}C^{(f)}$ that appears only
for $ N = 5 $, each of these eigenvalues is the sum of two terms containing $
N^2 $. They are: ({\it i}) a term  $\mu^{(f)} N^2 $ that may be interpreted as 
an "interaction" of $ N $ elements ("intrinsic partons") treated on the same 
footing, and ({\it ii}) another term

\vspace{-0.1cm}

%(A.13)
$$
\mu^{(f)}\left(\varepsilon^{(f)\,2} - 1\right)P_N^{2}\;\;{\rm with}\;\;P_N = 
\left[N!/(N-1)! \right]^{-1} = N^{-1} \eqno({\rm A}.13)
$$

\vspace{-0.1cm}

\ni that may describe an additional "interaction" with itself of one element 
arbitrarily chosen among~~$ N $~~elements of which the remaining~~$ N-1 $~~are 
undistinguish\-able. Therefore, the total "interaction" with itself of this 
(arbitrarily) distinguished "parton" is~~$\mu^{(f)}[1 + $ $ (\varepsilon^{(f)
\,2} - 1)N^{-2}]$, so it becomes $\mu^{(f)}\varepsilon^{(f)\,2}$ in the first 
fermion family.

The form (A.11) of three fermion wave functions shows that that each "intrinsic
parton" carries a Dirac bispinor index (of the Jacobi type). For the (arbitrar%
ily) distinguished "parton", this index, considered in the framework of a 
fermion wave equation, is coupled to the external gauge fields of the Standard 
Model. Thus, this "parton" carries the total spin 1/2 of the fermion as well as
a set of its Standard--Model charges corresponding to $f = \nu\,,\,e\,,\,u\,,
\,d $. For the $ N-1 $ undistinguishable "partons", obeying Fermi statistics 
along with the Pauli principle, their Dirac bispinor indices are mutually 
coupled, resulting into Lorentz scalars, while their number $ N-1 = 0,2,4 $ 
differen\-tiates between three fermion families (for each $ f = \nu\,,\,e\,,\,
u\,,\,d $). These "partons" are free of Standard--Model charges.

 Evidently, the intriguing question arises, how to interpret two possible boson
families corresponding to the number $N-1 = 1,3$ of undistinguishable "partons"
[10]. In the present paper this problem is not discussed. Here, we would like 
only to point out that three fermion families $ N = 1,3,5 $ differ from these 
two hypothetic boson families $ N = 2,4 $ by the full pairing of their $ N-1 = 
0,2,4 $ undistinguishable "partons". So, the boson families, containing an odd 
number $ N-1 = 1,3 $ of such "partons", might be considerably heavier. Note 
that the wave functions corresponding to $N = 2,4 $ can be reduced (under some 
relativistic requirements) to two other wave functions carrying only spin 0,

\vspace{-0.3cm}

%(A.14)
\begin{eqnarray*}
\phi^{(f)}_{2} & \equiv & \frac{1}{2\sqrt{2}}\left(C^{-1}\gamma^5 \right)_{
\alpha_1\alpha_2} \psi^{(f)}_{\alpha_1\alpha_2}  = \frac{1}{\sqrt{2}}\left( 
\psi^{(f)}_{12} - \psi^{(f)}_{21} \right) = \frac{1}{\sqrt{2}}\left( \psi^{
(f)}_{34} - \psi^{(f)}_{43}\right)\;, \\ 
\phi^{(f)}_{4} & \equiv & \frac{1}{6\sqrt{4}}\varepsilon_{\alpha_1\alpha_2
\alpha_3\alpha_4}\psi^{(f)}_{\alpha_1\alpha_2\alpha_3\alpha_4} = \frac{1}{
\sqrt{4}}\left( \psi^{(f)}_{1234} - \psi^{(f)}_{2134} + \psi^{(f)}_{3412} - 
\psi^{(f)}_{4312} \right)\;,
\end{eqnarray*} 

\vspace{-1.55cm}

\begin{flushright}
(A.14)
\end{flushright}

\vspace{-0.2cm}

\ni and appearing (up to the sign) with the multiplicities 2 and 6, 
respectively.

Another important question also appears, namely, what is the interpretation of 
two fermions corresponding to the number $N = 1\,,\,3 $  of undistinguishable 
"partons" only. Such fermions can carry exclusively spin 1/2 (for $N = 3 $: 
under some relativistic requirements). Of course, they are free of Standard--%
Model charges and so, can be considered as two sterile neutrinos with the wave 
functions

\vspace{-0.4cm}

%(A.15)
\begin{eqnarray*}
\!\!\nu_{s\,\alpha_1} & \equiv & \psi_{1\,\alpha_1} \equiv \psi_{\alpha_1}\;, 
\\ \!\!\nu'_{s\,\alpha_1} & \equiv & \psi_{3\,\alpha_1} \equiv \frac{1}{6}
\left(C^{-1}\gamma^5 \right)_{\alpha_1\,\alpha_2}\varepsilon_{\alpha_2\alpha_3
\alpha_4\alpha_5}\psi_{\alpha_3\alpha_4\alpha_5} = \left\{\begin{array}{r}
\psi_{134}\;{\rm for}\;\alpha_1 = 1 \\ -\psi_{234}\;{\rm for}\;\alpha_1
= 2 \\ \psi_{312}\;{\rm for}\;\alpha_1 = 3 \\ -\psi_{412}\;{\rm for}\;
\alpha_1 = 4 \end{array}\right.
\end{eqnarray*} 

\vspace{-2.05cm}

\begin{flushright}
(A.15)
\end{flushright}

\vspace{0.8cm}

\ni appearing (up to the sign) with the multiplicities 1 and 6, respectively.

 For these sterile neutrinos one may introduce the~~$2\times 2$~~mass matrix~~$
\widehat{M}^{(s)} = $ $\widehat\rho^{(s)\,1/2} \widehat{h}^{(s)} 
\widehat\rho^{(s)\,1/2}$, where

%(A.16)
$$
\widehat\rho^{(s)\,1/2} = \frac{1}{\sqrt{7}}\left(\begin{array}{cc} 1 & 0 \\
0 & \sqrt{6} \end{array} \right)\;\;,\;\; {\rm Tr}\widehat\rho^{(s)} = 1\;,
\eqno({\rm A}.16)
$$

\ni while the diagonal part of $\widehat{h}^{(s)}$ is conjectured to have the 
eigenvalues

%(A.17)
$$
\mu^{(s)}\left[N^2 + \left(\varepsilon^{(s)\,2} - 1\right)P^2_N\right]
\;\;{\rm with}\;\; P_N = N!/N! = 1\;\;(N = 1\,,\,3)\;.
\eqno({\rm A}.17)
$$

\ni Now, one "intrinsic parton" is arbitrarily chosen (to carry the total spin 
1/2 of the fermion) among $ N $ "intrinsic partons" that all are undistinguish%
able [in contrast to Eqs. (A.12) and (A.13)]. This gives the diagonal part of 
$\widehat{M}^{(s)} $ equal to

%(A.18)
$$
\frac{1}{7}\left(\begin{array}{cc} \mu^{(s)} \varepsilon^{(s)\,2} & 0 \\
0 & 6\mu^{(s)} (8+\varepsilon^{(s)\,2}) \end{array} \right)\;.
\eqno({\rm A}.18)
$$

\ni Thus, the diagonal matrix elements $M_{44}^{(\nu)}$ and $M_{55}^{(\nu)}$ 
of the $5\times 5$ neutrino mass matrix $\left(M_{IJ}^{(\nu)} \right)\;\;(I,J =
1,2,3,4,5)$  introduced in Eq. (11) get the forms

%(A.19)
$$
M_{44}^{(\nu)} = \frac{\mu^{(s)}}{7}\varepsilon^{(s)\,2} \simeq 0\;\;,\;\;
M_{55}^{(\nu)} = \frac{ 6\mu^{(s)}}{7}\left( 8+\varepsilon^{(s)\,2}\right) 
\simeq \frac{48\mu^{(s)}}{7} \eqno({\rm A}.19)
$$

\ni with $\varepsilon^{(s)\,2}$ expected to be very small. In the present paper
we will assume that 

%(A.20)
$$
\mu^{(s)} \sim \mu^{(\nu)}\;\;,\;\;\varepsilon^{(s)\,2} \sim 
\varepsilon^{(\nu)\,2}\;. \eqno({\rm A}.20)
$$

\ni in Eqs. (A.19).

 The possibility of existence of two bosons corresponding to the number $ N =2,
4 $ of undistinguishable "partons" only ought to be also considered. Such 
bosons can carry exclusively spin 0 (for $ N = 2 $: under some relativistic 
requirements). Obviously, they are free of Standard--Model charges and so, may 
be considered as two "sterile scalars" with the wave functions

%(A.21)
$$
\phi_2 \equiv \frac{1}{4}\left(C^{-1}\gamma^5 \right)_{\alpha_1\alpha_2}
\psi_{\alpha_1\alpha_2} = \psi_{12} = \psi_{34}\;,\;\phi_4 \equiv \frac{1}{24}
\varepsilon_{\alpha_1\alpha_2\alpha_3\alpha_4}\psi_{\alpha_1\alpha_2\alpha_3
\alpha_4} = \psi_{1234}  \eqno({\rm A}.21)
$$

\ni appearing (up to the sign) with the multiplicities 4 and 24, respectively.
  
 {\it A priori}, the "intrinsic partons" may be either strictly {\it algebraic}
objects providing fundamental fermions (leptons and quarks) with new family 
degrees of freedom, or may give us a signal of a new {\it spatial} substructure
of fundamental fermions (built up of spatial "intrinsic partons" = preons, 
related to the individual $\gamma^\mu_i $ as well as $ x^\mu_i $ and $ p^\mu_i
\;\;(i = 1,2,\ldots,N)$; note that here $\gamma^\mu_i\, $'s anticommute for 
different $ i $~!). Our idea about the origin of three fermion families
[4] chooses the first option. The difficult problem of new non--Standard--Model
forces, responsible for the binding of $N$ preons within fundamental fermions, 
does not arise in this option. 

 However, if the second option is true, then this irksome (though certainly 
profound) problem does arise and must be solved.

\vfill\eject

~~~~
\vspace{0.6cm}

{\bf References}

\vspace{1.0cm}

{\everypar={\hangindent=0.5truecm}
\parindent=0pt\frenchspacing

{\everypar={\hangindent=0.5truecm}
\parindent=0pt\frenchspacing

~1.~Y. Fukuda {\it et al.} (Super--Kamiokande Collaboration) , "Evidence for 
oscillation of atmospheric neutrinos", to appear in {\it Phys. Rev. Lett.};
and references therein.

\vspace{0.15cm}

~2.~M. Appolonio {\it et al.} (CHOOZ Collaboration), {\it Phys. Lett.} {\bf B 
420}, 397 (1998).

\vspace{0.15cm}

~3.~Y. Fukuda {\it et al.} (Kamiokande Collaboration), {\it Phys. Lett.} {\bf 
B 335}, 237 (1994).

\vspace{0.15cm}

~4.~W.~Kr\'{o}likowski, {\it Acta Phys. Pol.} {\bf B 21}, 871 (1990); {\it 
Phys. Rev.} {\bf D 45}, 3222 (1992); in {\it Spinors, Twistors, Clifford 
Algebras and Quantum Deformations (Proc. 2nd Max Born Symposium 1992)}, eds. 
Z.~Oziewicz {\it et al.}, 1993, Kluwer Acad. Press. 

\vspace{0.15cm}

~5.~W.~Kr\'{o}likowski, {\it Acta Phys. Pol.} {\bf B 27}, 2121 (1996); {\bf B 
28}, 1643 (1997); {\bf B 29}, 629 (1998); {\bf B 29}, 755 (1998); 
hep--ph/9803323.

\vspace{0.15cm}

~6.{\it ~Review of Particle Physics}, {\it Phys. Rev.} {\bf D 54}, 1 (1996), 
Part I. 

\vspace{0.15cm}

~7.~N.~Hata and P.~Langacker, hep--ph/9705339; {\it cf.} also G.L.~Fogli, 
E.~Lisi and D.~Montanino, hep--ph /9709473.

\vspace{0.15cm}

~8.~L.~Wolfenstein, {\it Phys. Rev.} {\bf D 17}, 2369 (1978); S.P. Mikheyev 
and A.~Y.~Smirnow, {\it Sov. J. Nucl. Phys.} {\bf 42}, 913 (1985); {\it 
Nuovo Cimento}, {\bf C 9}, 17 (1986).

\vspace{0.15cm}

~9.~C.~Athanassopoulos {\it et al.} (LSND Collaboration), {\it Phys. Rev.} 
{\bf C 54}, 2685 (1996); {\it Phys. Rev. Lett.} {\bf 77}, 3082 (1996);
nucl--ex/9709006.

\vspace{0.15cm}

10.~W.~Kr\'{o}likowski, {\it Phys. Rev.} {\bf D 46}, 5188 (1992); {\it Acta 
Phys. Pol.} {\bf B 24}, 1149 (1993); {\bf B 26}, 1217 (1995); {\it Nuovo 
Cimento }, {\bf 107 A}, 69 (1994).

\vfill\eject

\end{document}